\documentclass[runningheads]{llncs}
\usepackage{graphicx}
\usepackage{amssymb, amsmath, bm, latexsym,comment}
\usepackage{multirow}
\usepackage{xcolor}
\usepackage{subfigure}
\usepackage{indentfirst}
\usepackage{setspace}
\usepackage{verbatim}
\usepackage{array}
\usepackage{arydshln}
\usepackage[misc]{ifsym} 
\usepackage[colorlinks,linkcolor=blue]{hyperref}
\usepackage{xcolor}
\usepackage[normalem]{ulem} 

\providecommand{\Leireftb}[1]{Table~\ref{#1}}
\providecommand{\Leireffig}[1]{Fig.~\ref{#1}}

\providecommand{\citep}[1]{\cite{#1}}

\newcommand{\green}{\textcolor{green}}

\begin{document}
\title{Right Ventricular Segmentation from Short- and Long-Axis MRIs via Information Transition}
\titlerunning{SL-Segnet}

\author{ 
Lei Li \inst{1,2}\thanks{The two authors have equal contributions to the paper.}  \and
Wangbin Ding \inst{3}$^{\star}$ \and
Liqun Huang \inst{3} \and
Xiahai Zhuang\inst{1}${^{(\textrm{\Letter})}}$
} 
\authorrunning{L. Li et al.}

\institute{
School of Data Science, Fudan University, Shanghai, China \and
School of Biomedical Engineering, Shanghai Jiao Tong University, Shanghai, China \and
College of Physics and Information Engineering, Fuzhou University, Fuzhou, China \\
\email{zxh@fudan.edu.cn}
}

\maketitle 
\begin{abstract}
Right ventricular (RV) segmentation from magnetic resonance imaging (MRI) is a crucial step for cardiac morphology and function analysis.
However, automatic RV segmentation from MRI is still challenging, mainly due to the heterogeneous intensity, the complex variable shapes, and the unclear RV boundary.
Moreover, current methods for the RV segmentation tend to suffer from performance degradation at the basal and apical slices of MRI.
In this work, we propose an automatic RV segmentation framework, where the information from long-axis (LA) views is utilized to assist the segmentation of short-axis (SA) views via information transition.
Specifically, we employed the transformed segmentation from LA views as a prior information, to extract the ROI from SA views for better segmentation. 
The information transition aims to remove the surrounding ambiguous regions in the SA views. 
We tested our model on a public dataset with 360 multi-center, multi-vendor and multi-disease subjects that consist of both LA and SA MRIs.
Our experimental results show that including LA views can be effective to improve the accuracy of the SA segmentation.
Our model is publicly available at \url{https://github.com/NanYoMy/MMs-2}.
\end{abstract}

\section{Introduction}
The segmentation of right ventricular (RV) is an essential preprocessing step for the cardiac functional assessment, such as the volume of ventricles, regional wall thickness, and ejection fraction.
Manual delineations of the RV from short-axis (SA) and long-axis (LA) MRIs can be subjective and labor-intensive.
However, automatic RV segmentation remains challenging, mainly due to the heterogeneous intensity, the complex variable shapes, and the unclear boundary of RV \citep{journal/JTEHM/chen2018}.

In literature, most methods jointly segment both ventricles, and only a few methods focus exclusively on RV segmentation \citep{journal/JHE/zhuang2013,journal/MedIA/zhuang2020,journal/TMI/campello2021}. 
The joint segmentation of ventricles aims to employ the similar gray levels in their blood cavities and the relatively stable positions of two ventricles.
Therefore, conventional atlas-based methods and model-based approaches combining with prior anatomical knowledge, are commonly used in these joint optimization methods \citep{journal/MedIA/petitjean2015}.
Recently, with the development of deep learning (DL) in medical image computing, several DL-based algorithms have been proposed for automatic RV segmentation \citep{journal/CinC/luo2016,journal/MedIA/vigneault2018,journal/TBME/li2019}. 
The superiority of employing both the SA and LA images instead of only the SA images has been demonstrated \citep{conf/MICCAI/koikkalainen2004}. 
However, most current RV segmentation studies mainly focus on the algorithms solely using SA cardiac MRI \citep{journal/IET/ammari2021}.
Instead, the research on employing other views of MRIs to guide the segmentation of SA especially on the apical and basal slices, is rather rare.
Moreover, due to the scarce of multi-center and multi-disease clinical dataset, the challenges of RV segmentation on the data from different centers and pathologies are rarely considered.



In this work, we propose a multi-view (LA and SA view) segmentation framework to delineate RV from multi-center and multi-disease MRIs. 
The framework is consists of a 2D and a 3D nnU-Net \citep{journal/nature/isensee2021}, which aim to segment RV from LA and SA views in successive. 
The nnU-Net has the advantage of self-automatic configuration, and therefore alleviates the burden of manual effort in the network configuration. 
Moreover, LA views can provide comprehensive information for the apical and basal slices of SA views, and also visualize atria clearly. 
We therefore employ an information transition scheme to assist the SA view segmentation via the corresponding LA view. 

\subsubsection{Related Literature.}
For the literature of the RV segmentation, one could refer to the review paper \citep{journal/IET/ammari2021}, where over forty research papers were evaluated. 
The review paper showed that current RV segmentation methods still can not properly solve all existing RV challenging issues. 
For the simultaneous SA and LA MRI segmentation, Koikkalainen et al. \citep{conf/MICCAI/koikkalainen2004} employed both SA and LA MRI to segment ventricles and atria by transforming them into a same coordinate system.
Vigneault et al. \citep{journal/MedIA/vigneault2018} proposed an $\Omega$-net to segment ventricles and atria from MRIs with SA, four-chamber and two-chamber views.
They simultaneously transformed all these views into a canonical orientation, and then performed the segmentation on the transformed images.
Oghli et al. \citep{journal/PM/Oghli2018} assumed that RV cavity is continuous in the LA direction, and then transited the seed point of region growing method along the LA direction.
Chen et al. \citep{conf/MICCAI/chen2019} segmented left ventricular (LV) myocardium from SA views by combining the learned anatomical shape priors from various views.
It is still an open question about how to effectively employ LA views for the RV segmentation of SA views.

\section{Methodology}

\begin{figure*}[t]\center
 \includegraphics[width=0.98\textwidth]{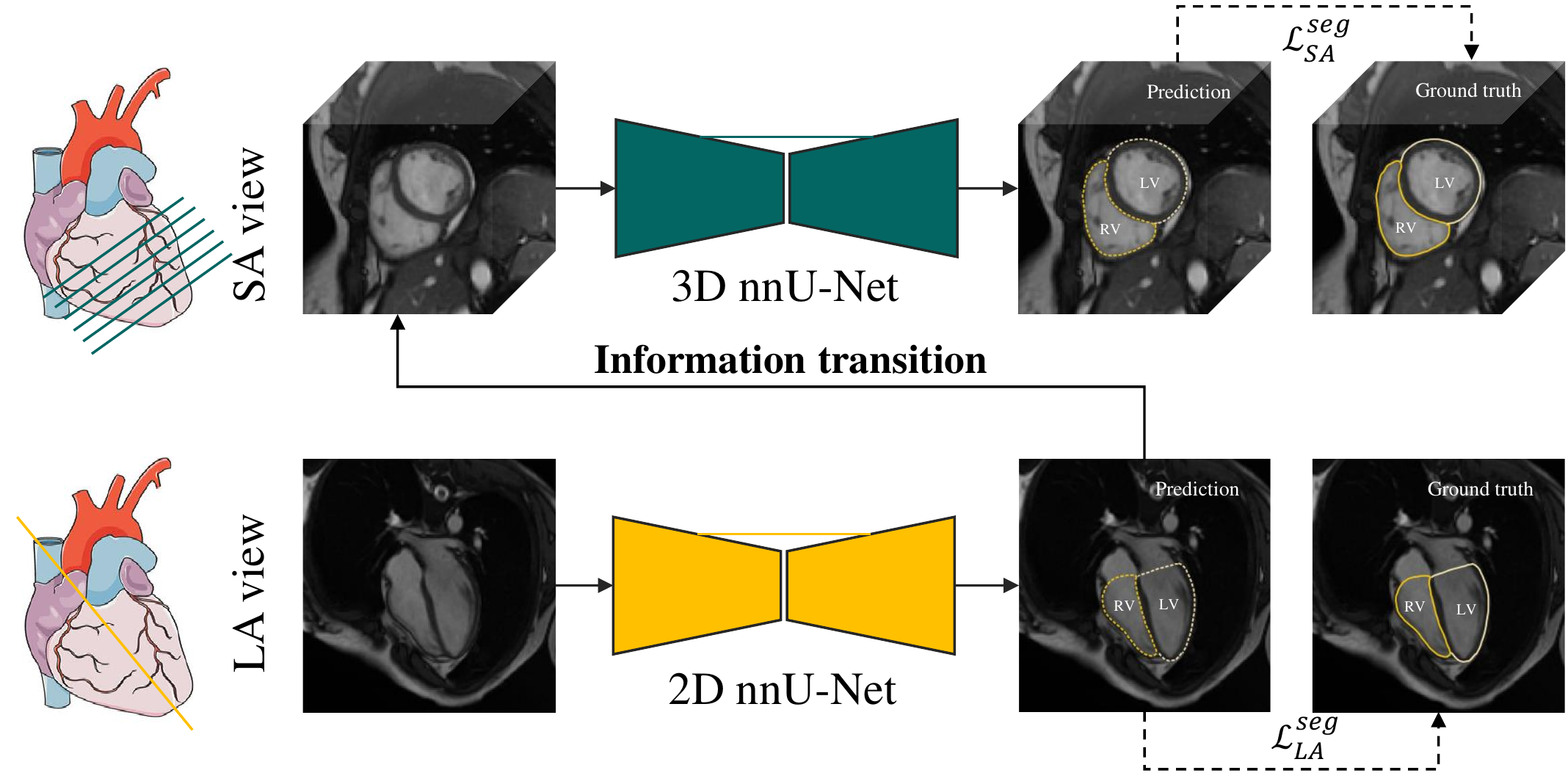}\\[-2ex]
   \caption{The proposed RV segmentation framework for both SA and LA images. 
   The framework includes three steps: the LA segmentation, ROI extraction from SA with assistant of LA information, and the SA segmentation.
   Here, the 3D cardiac image adopted from Kevil et al. \citep{journal/ATVB/Kevil2019}.}
\label{fig:method:framework}
\end{figure*}

\subsection{Segmentation Framework} \label{method:framework}
\Leireffig{fig:method:framework} presents the proposed segmentation framework, where SA and LA images are segmented via 3D and 2D nnU-Net \citep{journal/nature/isensee2021}, separately.
We firstly segment RV and LV from LA images, and then use this segmentation to localize the ventricles, which is used to guide the SA view segmentation.
More specifically, we transform this information into the coordinate system of SA view, and utilize this information to crop the SA view (see Section \ref{method:information transition}).
The segmentation loss functions of the framework are defined as follows,
\begin{equation}
 \mathcal{L}_{SA}^{seg}=\mathcal{L}_{SA}^{CE}+\lambda_{SA} \mathcal{L}_{SA}^{Dice},
 \label{eq:seg_sa}
\end{equation}
\begin{equation}
 \mathcal{L}_{LA}^{seg}=\mathcal{L}_{LA}^{CE}+\lambda_{LA} \mathcal{L}_{LA}^{Dice},
\label{eq:seg_la}
\end{equation}
where $\lambda_{SA}$ and $\lambda_{LA}$ are balancing parameters, and $\mathcal{L}^{CE}$ and $\mathcal{L}^{Dice}$ are the cross entropy (CE) loss and Dice loss, separately.
Note that though our final target is to segment RV, here we also include the LV (both LV cavity and myocardium) label when minimizing the loss.
We argue that the relatively stable space relationship of two ventricles can be helpful for the RV segmentation, especially in the boundary regions.
Besides, we do not separate the LV cavity and myocardium to avoid overly attention on the supervision of noncritical small targets, i.e., LV myocardium.

\subsection{Information Transition from the LA view} \label{method:information transition}

\begin{figure*}[t]\center
 \includegraphics[width=0.98\textwidth]{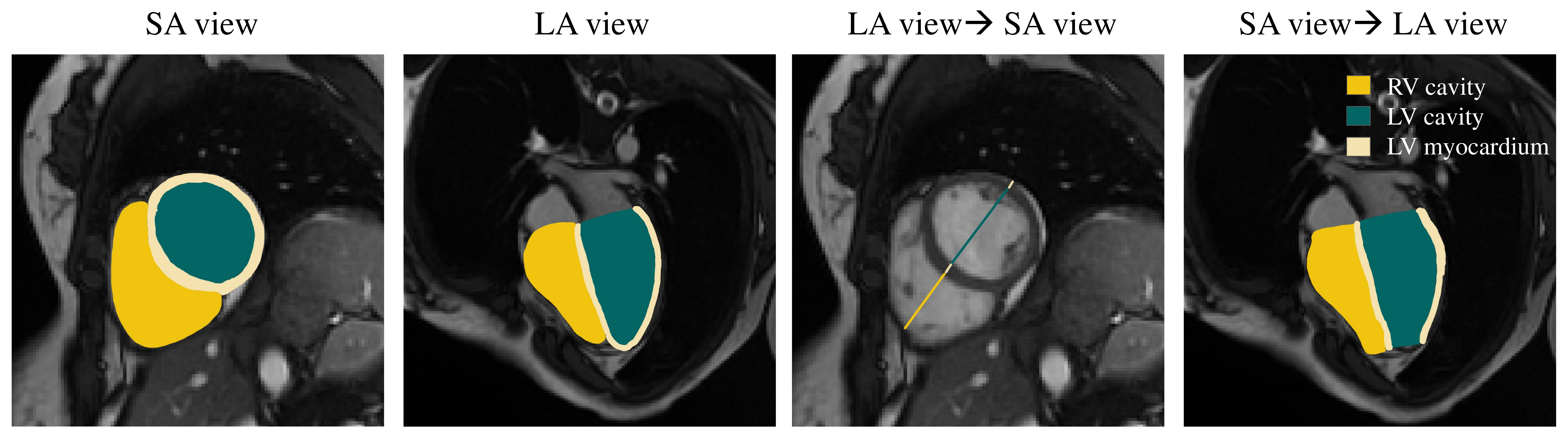}\\[-2ex]
  \caption{Example of transformations between the the label images of the SA and LA views.
  One can see that the transformed LA label only has one straight line traversing the ventricles.
  In contrast, the transformed SA label covers the whole ventricles but missing some apical regions.}
\label{fig:method:transform}
\end{figure*}

To employ the information from the LA view, we need to align the SA and LA views into  a common coordinate system, as shown in \Leireffig{fig:method:transform}.
The transformation parameter between SA and LA views can be extracted from the header information of images.
Specifically, the physical coordinates of SA and LA views can be defined as follows,
\begin{equation}
x^{\prime}_{SA}=T_{SA}(x_{SA}), 
\end{equation}
\begin{equation}
x^{\prime}_{LA}=T_{LA}(x_{LA}), 
\end{equation}
where $T$ is the transformation matrix that converts the image coordinate $x$ into the physical coordinate $x^{\prime}$.
We assume that the physical coordinates of SA and LA views are consist, so the transformed LA and SA views can be defined as follows,
\begin{equation}
x_{LA\rightarrow SA}=T_{SA}^{-1}(T_{LA}(x_{LA})),
\end{equation}
\begin{equation}
x_{SA\rightarrow LA}=T_{LA}^{-1}(T_{SA}(x_{SA})).
\end{equation}


\begin{figure*}
  \center
 \includegraphics[width=0.65\textwidth]{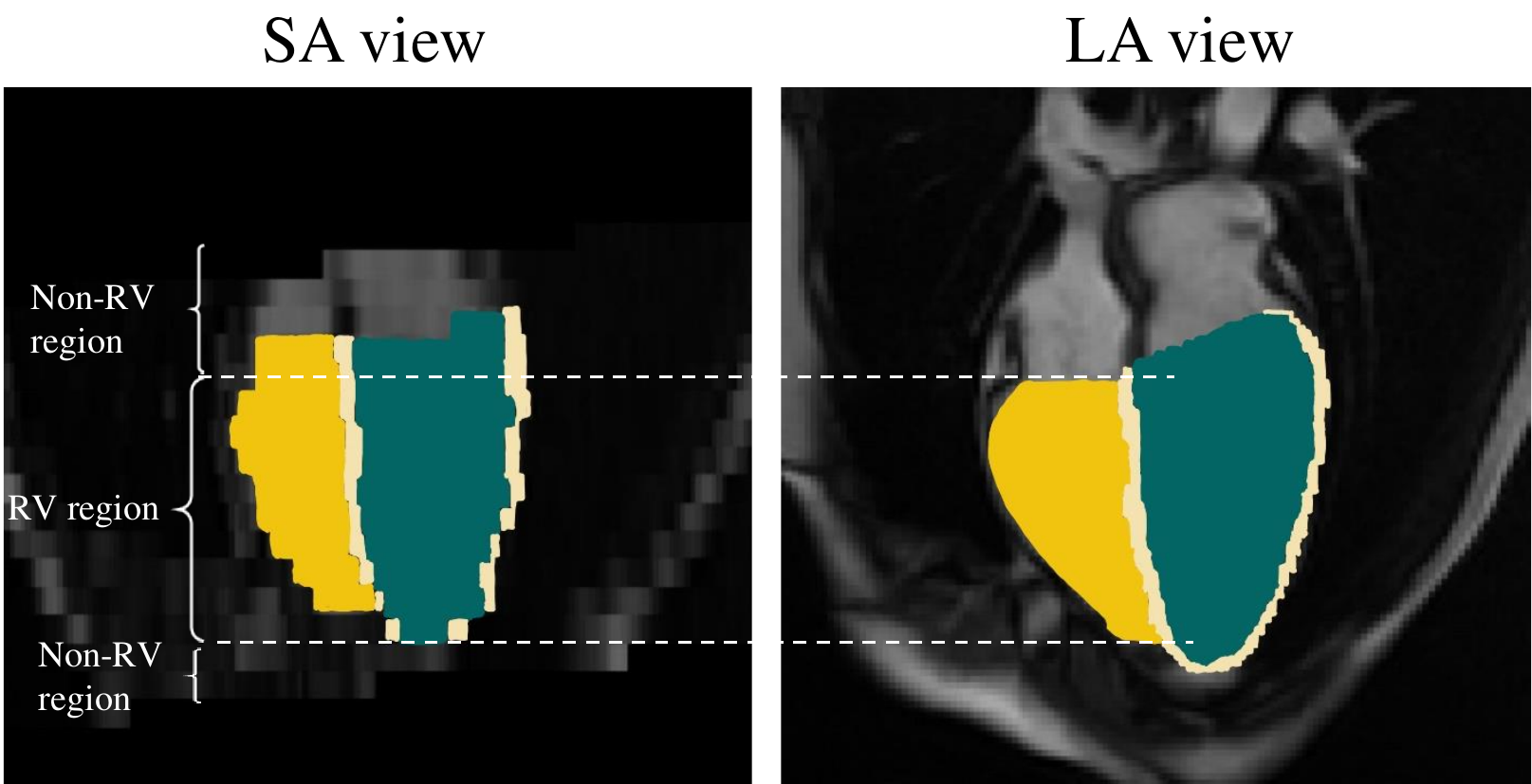}\\[-2ex]
    \caption{The correspondence between SA and LA views in the coronal planes. 
    There may be some inconsistencies in the apical and basal regions between the SA and LA views.
    Therefore, the SA plane can be marked as RV or non-RV region based on its correspondence with the LA plane.
    Note that here the orientation of images has been adjusted for better visualization.
     }
    \label{fig:exp:RV slice}
\end{figure*}

\Leireffig{fig:exp:RV slice} presents the aligned LA and SA views in the coronal plane.
One can see that in the SA view, the basal regions of the RV tends to be confused with the right atrium.
In contrast, LA views can provide relatively clear boundary in the ambiguous regions. 
Therefore, with the assist of the LA information, one can classify the SA view as RV or non-RV regions in the coronal plane.
Specifically, we employed the transformed LA segmentation (see \Leireffig{fig:method:transform}) as a prior information, to extract the ROI from SA views for better segmentation.
\textit{Note that, the ROI excludes the aforementioned non-RV regions, where the SA segmentation tends to be inaccurate.}

\section{Materials}

\subsection{Data Acquisition and Pre-processing}
The dataset is from the \textit{Multi-Disease, Multi-View \& Multi-Center Right Ventricular Segmentation in Cardiac MRI (M\&Ms-2)} \citep{link/MMs2021} challenge event.
The challenge dataset is consisted of 360 multi-center and multi-vendor subjects that are divided into three parts: 160 training data, 40 validation data, and 160 test data.
It covers both healthy volunteers and patients with different pathologies in both SA and LA views, as presented in \Leireftb{tb:data info}. 
Two pathologies (tricuspidal regurgitation and congenital arrhythmogenesis) do not appear in the training dataset, but are included in the validation and testing sets.
The data setting aims to evaluate the model generalization ability to unseen pathologies.

\begin{table*} [t] \center
    \caption{
    Pathology distribution among the training data, validation data, and test data.
    HCM: hypertrophic cardiomyopathy; CAM: congenital arrhythmogenesis; TOF: tetralogy of fallot;
    IC: interatrial comunication; TR: tricuspidal regurgitation.
     }
\label{tb:data info}
{\small
\begin{tabular}{  l| l l l *{4}{@{\ \,} l }}
\hline
Pathology & Num. training & Num. validation & Num. test \\
\hline
Normal/Dilated LV/HCM  & 40/30/30 & 5/5/5 & 30/25/25 \\
CAM/TOF/IC             & 20/20/20 & 5/5/5 & 10/10/10 \\
Dilated RV/TR          & 0/0      & 5/5   & 25/25 \\
\hline
\end{tabular} }\\
\end{table*}

\subsection{Gold Standard and Evaluation}
All the MRIs were manually delineated by experienced clinicians from the respective centers, and the label consistency between SA and LA images in basal and apical slices was confirmed.
The manual segmentation includes the contours of RV, LV and LV myocardium. 
As this study focus on the RV segmentation, we only employ the RV manual label as the gold standard in the final evaluation.

For evaluation, Dice score (DS) and Hausdorff distance (HD) were applied. 
The final evaluation score is obtained by applying 0.75 and 0.25 weighting coefficients to the SA and LA segmentation accuracy, respectively. 
\begin{equation}
    \mathrm{score}=\frac{0.75(\text{DS}_{SA} + \text{HD}_{SA}) + 0.25(\text{DS}_{LA} + \text{HD}_{LA})}{2},
\end{equation}
where $\text{DS}_{SA/LA} = (\text{DS}_{ED} + \text{DS}_{ES})/2$ and $\text{HD}_{SA/LA} = (\text{HD}_{ED} + \text{HD}_{ES})/2$.

\subsection{Implementation} 
The proposed framework was implemented in PyTorch, running on a computer with a Core i7 CPU and an NVIDIA GeForce RTX 1080. 
To train the segmentation networks in proposed framework, $\lambda_{SA}$ and $\lambda_{LA}$ were set to 1 (see Equation \ref{eq:seg_sa} and \ref{eq:seg_la}). 
An Adam optimizer with an initial learning rate of 0.01 was adopted, and the networks were trained with 500 epochs. 

\section{Experiment}

\subsection{Comparison Experiment} 
We implemented a baseline scheme and three utilization strategies to employ LA information for the segmentation of SA views.
\begin{itemize}
    \item[(1)] \textbf{W/o-utilization}: one can train a nnU-Net purely on SA views without using any information from LA views. It can be considered as the baseline. 
    \item[(2)] \textbf{Post-utilization}: one can remove the non-RV regions of SA views via the prior segmentation of LA views. 
    \item[(3)] \textbf{Joint-utilization}: one can train a modified nnU-Net with an additional slice-level task at the bottom of network \citep{conf/MICCAI/yue2019}. Here, the additional task aims to identify whether a slice includes the RV regions. 
    \item[(4)] \textbf{Pre-utilization}: The proposed framework. One can first perform ROI extractions on SA views via transformed LA views (see Section \ref{method:information transition}), and then train a nnU-Net to segment the RV on the ROI. 
\end{itemize}

\begin{table*} [t] \center
    \caption{
    The performance on the validation set using different schemes to utilize LA information for the RV segmentation of SA views. 
    The best and second results are in \textbf{bold} and \underline{underline}, respectively.
     }
\label{tb:exp:comparison}
{\small
\begin{tabular}{ l| l *{4}{@{\ \,} l }}\hline
Matrix       & W/o-utilization & Post-utilization  & Joint-utilization & Pre-utilization \\
\hline
DS$_{SA}$ $\uparrow$        &\quad $ \textbf{0.914} $&  \quad $ 0.901 $&  \quad $ 0.900 $&  \quad \underline{$ 0.913 $}  \\
HD$_{SA}$ (mm) $\downarrow$ &\quad $ 11.2  $&  \quad $ 11.3  $&  \quad $\textbf{ 10.5}$&  \quad \underline{$ 10.6  $}  \\
\hline
\end{tabular} }\\
\end{table*}

\Leireftb{tb:exp:comparison} presents the results of different strategies on validation dataset. 
Though w/o-utilization strategy obtained the best DS, it performed slightly worse than the joint- and pre-utilization schemes in terms of HD. 
The post-utilization scheme did not present any advantages compared to the baseline, and the joint-utilization strategy tended to decrease the DS.
Therefore, we argue that the pre-utilization scheme is the most reliable and robust among all these strategies.

\subsection{Performance on the Data with Different Pathologies} 

\Leireftb{tb:exp:pathology} presents the accuracy of each pathology on the test data.
One can see that the best performance was obtained on the subjects with congenital arrhythmogenesis (CAM), though the most number of training data is from normal subjects.
It may indicate that the accuracy of each pathology did not solely rely on the number of training data.
There are two unseen pathologies in the training stage (see \Leireftb{tb:data info}), i.e., dilated RV and TR.
One can see that only the accuracy of dilated RV had an evident decrease for the segmentation of LA views.
It can be attribute to the irregular RV shape in the dilated RV patients.
Therefore, the model generalization ability on the unseen pathologies is generally promising.

\begin{table*} [t] \center
    \caption{
    The performance on the test data for each pathology. Here, \green{$^{\dagger}$} denotes the unseen pathologies in the training stage.
     }
\label{tb:exp:pathology}
{\small
\begin{tabular}{  l| l l l l *{5}{@{\ \,} l }}
\hline
Pathology & DS$_{SA}$ & \quad HD$_{SA}$ (mm) & \quad DS$_{LA}$ & \quad HD$_{LA}$ (mm) \\
\hline
Normal      & $ 0.916 \pm 0.042 $ & \quad $ 9.36 \pm 4.14 $ &\quad $ 0.931 \pm 0.033 $ &\quad $ 5.27 \pm 3.00 $ \\
Dilated LV  & $ 0.920 \pm 0.069 $ & \quad $ 11.2 \pm 5.53 $ &\quad $ 0.915 \pm 0.052 $ &\quad $ 6.08 \pm 3.13 $\\
HCM         & $ 0.930 \pm 0.052 $ & \quad $ 9.11 \pm 4.86 $ &\quad $ 0.926 \pm 0.033 $ &\quad $ 5.35 \pm 2.62 $\\
CAM         & $ 0.943 \pm 0.025 $ & \quad $ 8.52 \pm 4.04 $ &\quad $ 0.934 \pm 0.034 $ &\quad $ 5.88 \pm 5.70 $ \\
TOF         & $ 0.920 \pm 0.035 $ & \quad $ 11.8 \pm 2.72 $ &\quad $ 0.909 \pm 0.034 $ &\quad $ 7.56 \pm 3.16 $ \\
IC          & $ 0.915 \pm 0.042 $ & \quad $ 11.6 \pm 4.18 $ &\quad $ 0.916 \pm 0.070 $ &\quad $ 6.43 \pm 4.49 $\\
Dilated RV\green{$^{\dagger}$}  & $ 0.917 \pm 0.047 $ & \quad $ 11.1 \pm 3.30 $ &\quad $ 0.888 \pm 0.136 $ &\quad $ 7.80 \pm 8.32 $ \\
TR\green{$^{\dagger}$}          & $ 0.910 \pm 0.047 $ & \quad $ 10.5 \pm 5.52 $ &\quad $ 0.915 \pm 0.036 $ &\quad $ 5.98 \pm 3.05 $\\
\hline
\end{tabular} }\\
\end{table*}


\subsection{Performance on the ED and ES Phase} 
\Leireftb{tb:exp:phase} presents the quantitative results of the proposed method on the ED and ES phases.
One can see that the performance on the ES phase was statistically significant ($p\textless 0.001$) worse than that on the ED phase in terms of DS, but no significant difference ($p\textgreater0.1$) in terms of HD.
As we know Dice score belongs to volumetric overlap measurement, and can be sensible to the size of target volume.
Therefore, it could attribute to the larger surface of RV in the ED phase compared to that in the ES phase.

\begin{table*} [t] \center
    \caption{
    The performance on the test data for both ED and ES phases. 
     }
\label{tb:exp:phase}
{\small
\begin{tabular}{l| l l l l *{5}{@{\ \,} l }}
\hline
Phase & DS$_{SA}$ & \quad HD$_{SA}$ (mm) & \quad DS$_{LA}$ & \quad HD$_{LA}$ (mm) \\
\hline
ED      & $ 0.933 \pm 0.039 $ & \quad $ 10.6 \pm 4.89 $ & \quad $ 0.930 \pm 0.050 $ & \quad $ 6.25 \pm 3.73 $ \\
ES      & $ 0.907 \pm 0.056 $ & \quad $ 10.1 \pm 4.45 $ & \quad $ 0.902 \pm 0.080 $ & \quad $ 6.10 \pm 5.38 $\\
\hline \hline
Average & $ 0.920 \pm 0.050 $ & \quad $ 10.3 \pm 4.67 $ & \quad $ 0.916 \pm 0.068 $ & \quad $ 6.17 \pm 4.61 $\\
\hline
\end{tabular} }\\
\end{table*}

\section{Conclusion}
In this work, we have proposed a framework for the RV segmentation of both SA and LA views.
The proposed model has been tested on 160 subjects and obtained promising results, even on the unknown pathologies.
The experimental results also demonstrated the effectiveness of the proposed information transition scheme.
A limitation of this work is that the SA and LA view segmentation are achieved separately, as LA segmentation is regarded as a prior for the SA segmentation.
In the future, we will develop more elegant and effective information transition algorithm for the simultaneous segmentation of SA and LA views.

\subsubsection{Acknowledgement.}
This work was funded by the National Natural Science Foundation of China (grant no. 61971142, 62111530195 and 62011540404) and the development fund for Shanghai talents (no. 2020015).

\bibliographystyle{splncs04}
\bibliography{A_refs}

\end{document}